\documentclass[prl,twocolumn,floatfix,superscriptaddress]{revtex4}

\usepackage[dvips]{graphicx}
\usepackage{epsfig}

\usepackage{verbatim}

\usepackage{times}

\usepackage{subfigure}

\usepackage{amsmath}

\usepackage{natbib}

\begin{document}

\title{Exponential distributions of collective flow-event properties in viscous
  liquid dynamics}

\newcommand{\angleb}[1]{\langle #1 \rangle}
\newcommand{\nod}{\noindent}

\date{\today}

%\pacs{}

\author{Nicholas P. Bailey}
\email{nbailey@ruc.dk}
\affiliation{Department of Sciences, DNRF Center 
``Glass and Time'', Roskilde University, P.O. Box 260, DK-4000 Roskilde, 
Denmark}

\author{Thomas B. Schr{\o}der}
\affiliation{Department of Sciences, DNRF Center 
``Glass and Time'', Roskilde University, P.O. Box 260, DK-4000 Roskilde, 
Denmark}

\author{ Jeppe C. Dyre }
\affiliation{Department of Sciences, DNRF Center 
``Glass and Time'', Roskilde University, P.O. Box 260, DK-4000 Roskilde, 
Denmark}

\begin{abstract} % max 600 characters incl spaces
We study the statistics of flow events in the inherent dynamics in
supercooled two- and three-dimensional binary Lennard-Jones liquids. 
Distributions of changes of the collective quantities 
energy, pressure and shear stress become exponential at low 
temperatures, as does that of the event ``size'' $S\equiv\sum {d_i}^2$.
We show how the $S$-distribution controls the others, while itself following
from exponential
tails in the distributions of (1) single particle displacements $d$, involving 
a Lindemann-like length $d_L$ and (2) the number of active particles 
(with $d>d_L$).
\end{abstract}

\maketitle

Many complex systems, including turbulent flows \cite{Pope:1993}, 
plastically deforming 
crystals \cite{Dimiduk/others:2006}, and financial markets \cite{Stanley:2003},
are characterized by intermittent, stochastic dynamics. It is important
to characterize the statistics of the
individual dynamical events, called ``increments'' in formal stochastic models,
``returns'' in finance, and ``avalanches'' in driven systems 
\cite{Bailey/others:2007}. In particular
non-Gaussian effects can have crucial significance, as
extreme events can dominate the dynamics. For example, in finance the fact that
distributions of returns are known to have ``fat'' (i.e. power-law) tails---in
opposition to assumptions made by standard models 
\cite{Black/Scholes:1973}---has been proposed as part of the reason for the 
failure to understand market crashes \cite{Bouchaud:2008}. In Gaussian 
statistics, values more than five standard deviations
from the mean account for less than one millionth of the distribution. For a
symmetric exponential distribution, on the other hand, they account for about
one thousandth. For the fat-tailed case of a Cauchy distribution 
$\gamma/\pi(\gamma^2+x^2)$, the equivalent
fraction (in terms of the standard deviation of the best-fit Gaussian)
is nearly one tenth. Examples of exponential distributions (which may
be considered intermediate between the Gaussian and power-law cases) include
bursts of protein production from gene expression \cite{Cai/Friedman/Xie:2006}
 and even returns in financial time series (at least 
for not-too large values) \cite{McCauley/Gunaratne:2003}.

% Both string-like \cite{Donati/others:1999,
% Schroder/others:2000} and larger, but more compact 
% (``democratic'') \cite{Appignanesi/others:2006} spatial
% correlations of particle displacements have been found.

Rearrangements of molecules known as ``flow events'' are the fundamental
processes giving rise to structural relaxation and flow in disordered systems
ranging from the equilibrium case of highly viscous 
liquids \cite{Goldstein:1969, Schroder/others:2000, Heuer:2008} to 
non-equilibrium systems such as glasses \cite{Bailey/others:2007}, granular 
materials \cite{Miller/OHern/Behringer:1996} and foams 
\cite{Pratt/Dennin:2003}. For viscous
(or ``glass-forming'') liquids, it has been accepted since Goldstein's 
1969 paper \cite{Goldstein:1969} that the dynamics may be understood
in terms of a division into fast vibrational motion around a 
particular energy minimum, and relatively rare transitions between neighboring
minima. It is the latter that are responsible for the slow 
dynamics \cite{Schroder/others:2000,Buchner/Heuer:2000}, thus a
detailed understanding of their nature is essential, particularly since many 
theoretical approaches start by making assumptions about the flow 
events \cite{Adam/Gibbs:1965,Dyre/Olsen/Christensen:1996,
Lubchenko/Wolynes:2007}.
Computer simulations provide the means to go beyond assumptions, and in recent
 years have provided much insight into the kinds of molecular motions that 
occur in a supercooled liquid.
Most workers have concentrated on particle displacements. But 
of more direct relevance to experiments are collective properties of the 
dynamics---changes in quantities such as potential energy $E$, pressure $p$,
and shear stress $\sigma_s$. Knowledge of their distributions, and how
those depend on temperature, is sure to be relevant for understanding the
slowing down of dynamics in viscous liquids.
Vogel et al. \cite{Vogel/others:2004} studied the
relation between particle displacements and the size of energy changes during
flow events, and while they reported exponential tails in both, they did not 
examine their relationship in detail, while Schr{\o}der et al. also
 reported such tails in flow event displacements \cite{Schroder/others:2000}.
Exponential distributions of forces in a simulated Lennard-Jones fluid were 
observed already in 1987 \cite{Powles/Fowler:1987}, although their relevance to
flow event properties is unclear.
It is now becoming possible to study individual particle displacements
experimentally, as Schall et al have done for a colloidal system 
\cite{Schall/Weitz/Spaepen:2007}; as yet the analysis is limited to a few
 events. The exponential tails recently reported for the van Hove correlation 
function \cite{Chaudhuri/Berthier/Kob:2007} reflect the 
cumulative effect of many flow events, whereas this work focuses on 
statistics on single flow events.

\begin{figure}
\epsfig{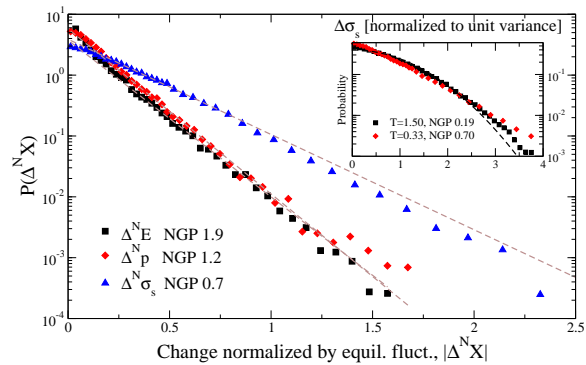}
\caption{\label{histEventSizes} (Color online) Distributions of 
$\Delta^N E$, $\Delta^N p$, $\Delta^N \sigma_{s}$ for $T=0.33$ (2D); 
dashed lines 
show exponential fits (without transforming by taking logarithms) to values
above 0.3, with decay lengths 0.160, 0.162, 0.284 for $E$, $p$,
 $\sigma_{s}$, respectively. Inset, distribution of $\Delta \sigma_s$ at
$T$=1.5 and 0.33, now normalized to have unit standard deviation, and
compared with a standard normal distribution. Non-Gaussian parameters (NGP) are
 indicated for each distribution.}
 \end{figure}

%% Summary of paper's result

In this Letter we present simulation data from two- and three-dimensional 
(2D and 3D) binary Lennard-Jones (BLJ) model fluids, brought to a viscous
state by cooling. By
studying the so-called ``inherent dynamics'' \cite{Schroder/others:2000}---the
 trajectory obtained by mapping configurations to local energy 
minima---we identify flow events and study their statistics.
Our main results are (1) at low temperatures $T$ the distributions of changes 
of $E$, $p$ and $\sigma_s$ are exponential, in contrast to high temperatures
where they are basically Gaussian; (2) the sum of squared 
displacements $S$, a geometrical measure of the size of an event,
 is the controlling quantity in the sense that it also has an 
exponential distribution (ED) at low temperatures, but for fixed $S$ values 
$E$, $p$ and $\sigma_s$ have Gaussian distributions. The mean event size
decreases with decreasing temperature; (3) the ED of $S$ can be 
traced to the existence of an exponential tail in the distribution of particle
 displacements during events, characterized by a Lindemann-like length 
scale, which defines ``active'' particles. The number of active particles 
is broadly distributed at high $T$, typically comprising a large 
fraction of the particles in the system, whereas at low $T$ it is
also exponentially distributed, with a mean of a few tens of particles.
This crossover coincides with the increasing relevance of minima
transitions to the dynamics at lower temperature. Indeed, at high temperature a
transition occurs almost every time step and their relevance to dynamics is
clearly minimal, whereas at lower temperatures  events occur one at a time in a
localized part of the sample. We leave analysis of the waiting times and 
correlations between events to later work.

The parameters for the BLJ potential are (where  $\epsilon$ and $\sigma$ are 
the energy and length scales for interactions between large (L-) and small
(S-) particles) $\epsilon_{LL}=1$, $\epsilon_{SS}=0.5$, 
$\epsilon_{LS}=1.5$, $\sigma_{LL}=1$, $\sigma_{SS}=0.88$,
$\sigma_{LS}=0.8$. All particles have the same mass $m=1$. These
parameters are identical to those of the BLJ
introduced by Kob and Andersen \cite{Kob/Andersen:1994}. The potential was
truncated using an interpolating polynomial between 2.4 $\sigma_{\alpha\beta}$ 
and  2.7 $\sigma_{\alpha\beta}$ ($\alpha,\beta \in \{L,S\}$). All results reported
here are from constant volume
simulations with periodic boundary conditions, with $N_p$=700 (1372) particles 
in 2D (3D), of which 60\% (80\%) were of type L, at a density of 
1.2$\sigma_{LL}^{-2}$ (1.2$\sigma_{LL}^{-3}$), using a time step
 of 0.01 $\sigma_{LL}\sqrt{m/\epsilon_{LL}}$ (0.005 at $T\ge$1.0). The system 
size was chosen to allow a large range of event sizes.
From now on, all quantities will be reported in the ``natural'' units defined
 by $\epsilon_{LL}$, $\sigma_{LL}$ and $m$.  Two-step relaxation, a signature of
 landscape-influenced dynamics, first appears at $T\sim0.50$ (2D) and $T=0.60$
(3D).

\begin{figure}
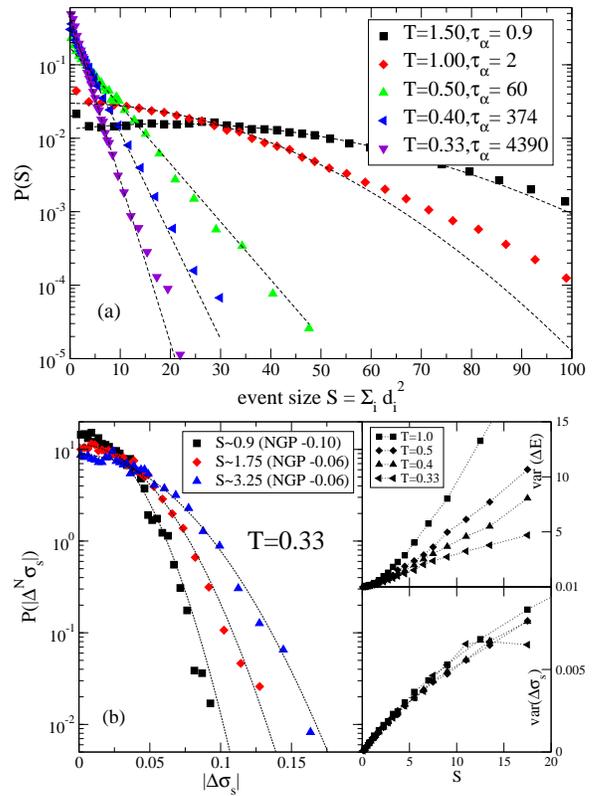

\epsfig{file=Sdistributions2.eps, width=3.0 in,clip=}

\epsfig{file=deltaShearFixedS2.eps, width=3.0 in,clip=}

\caption{\label{Sdistributions} (Color online) (a) Distributions of 
 $S$ for events for different temperatures, and
fits to the tails (Gaussian for T=1.0, 1.5, otherwise exponential). Structural
relaxation times $\tau_\alpha$ determined from the self-intermediate 
scattering function are also indicated. 
(b) Left panel, normalized change in shear stress for event size $S$ in
 different narrow intervals at $T=0.33$. Right panels, variance of 
(absolute) changes of (top) $E$ and (bottom) $\sigma_{s}$ as a function of 
$S$. For the range of $S$ relevant for
lower $T$, the variances are almost linear in $S$. This linear relation is 
independent of $T$ for $\sigma_{s}$, while the slope increases with $T$ for $E$.}
\end{figure}

To identify flow events we carried out Stillinger's
procedure \cite{Stillinger:1995} of quenching configurations to the 
``nearest'' local minimum. For $T\ge0.5$, we quenched every time 
step. For lower $T$ we quenched 
every tenth time step, and if a change of minimum was observed, the simulation
was ``backtracked'' and quenched at each of the intervening time steps. 
We detected events using the sum of squared particle displacements
$S=\sum_i d_i^2$, where $d_i$ is the magnitude of the displacement of the $i$th 
particle between successive inherent structures: A value greater than 10$^{-3}$ 
is sufficient to distinguish a genuine
change of minimum from numerical noise. This criterion is consistent with one
based on changes in inherent energy or stress. Care was exercised
in the minimization process \footnote{We used a combination of the 
``molecular dynamics'' method (MDmin) \cite{Stoltze:1997}
 and the conjugate gradient (CG) method \cite{Press/others:1987}.
MDmin follows the steepest descent path
reasonably closely, but frequent zeroing of velocities was needed  to avoid
``jumping a ridge'' and finding a different minimum (similarly if CG was 
started too soon). Spurious events, evident as pairs of 
equal and opposite events one time step apart, could not be entirely avoided
without prohibitive cost, but were removed from the analysis. Despite
the sensitivity of event detection to the minimization
procedure, the statistical properties (distributions) are quite robust.}.
In the following we place somewhat more emphasis on the 2D data because better 
statistics were obtained; due 
to a larger number of particles in 3D (though the linear size
is more than a factor of two smaller) and the larger number of neighbors 
per particle, the minimization is more time consuming (by about a factor of 
six). Thus 10--20 $\times 10^3$ events per temperature were obtained in 3D
compared to $10^5$ in 2D.

\begin{figure}
\epsfig{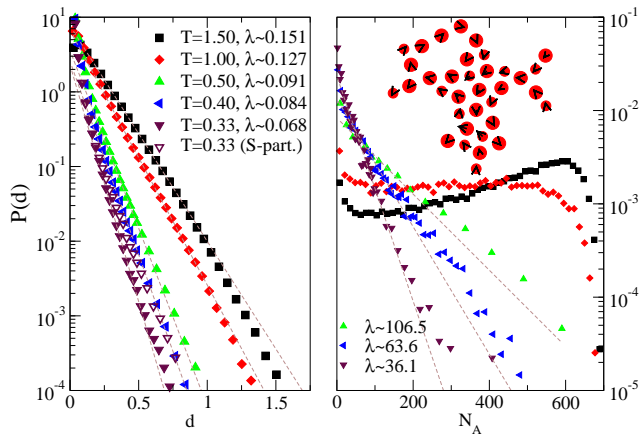}
\caption{\label{histFlowEventDisps} (Color online) Left panel, distributions of
 L-particle 
displacements during events at four temperatures, and exponential fits of
tails ($d>$0.3), with $\lambda$ denoting decay length. S-particle 
displacements are shown for the lowest temperature (where the difference is 
greatest). Right 
panel, distribution of $N_A$, the number of active particles, for the same
 temperatures, with fits for the three lowest temperatures. A visualization of
the 35 active particles in an event at $T$=0.33 is shown in the upper part.}
\end{figure}

Figure~\ref{histEventSizes} shows $P(\Delta^NX)$
as a function of $|\Delta^N X|$, where $X$ is $E$, $p$ 
or $\sigma_{s}$ and the $\Delta^N$ denotes the change normalized by the r.m.s.
equilibrium fluctuations ($\sqrt{\langle(X - \langle X \rangle)^2\rangle}$) 
calculated from the time series of inherent states ($T$=0.33).
The tails are very
close to exponential for $|\Delta^N X| \gtrsim 0.25$.
Though the distributions differ at small values, the
tails for $p$ and $E$ are very similar, with the same decay 
length $\lambda\sim$ 0.16 (possibly a reflection of the strong-pressure 
correlations recently reported for Van der Waals systems 
\cite{Pedersen/others:2008}), while that for shear is larger, 
$\lambda\sim$ 0.28. To quantify how close to Gaussian a 
distribution is we use a non-Gaussian parameter NGP$\equiv 
\langle(\delta x)^4\rangle/3(\langle(\delta x)^2\rangle)^2 - 1$ for any quantity
$x$. This is zero for a Gaussian distribution and unity for a (symmetric) ED.
The inset of the figure compares distributions of shear stress changes 
at the highest and lowest temperatures, now normalized to unit width. That from
$T$=1.5 is clearly more Gaussian, albeit with an exponential tail a few
 standard deviations from the mean.

By symmetry, changes of shear stress are uncorrelated with those of energy
 and pressure, meaning that none of these quantities can be considered a 
meaningful measure of the ``size'' of an event. A natural
measure of the size of an event is the quantity used to detect them, 
$S=\sum_i d_i^2$. The $S$ distributions are shown in 
Fig.~\ref{Sdistributions}(a). Going from high to low temperature, there is a
dramatic reduction in the average size, while the shape of the distribution
becomes increasingly exponential \footnote{Note that the 
dynamics at the highest temperatures cannot be activated---inherent quantities 
decorrelate as fast as real quantities. The data is shown
mainly for comparison.}. Examining the distributions of $\Delta X$ 
for a narrow range of $S$, we find they are Gaussian even at low $T$ when the 
full distributions are exponential (Fig.~\ref{Sdistributions}(b)).
For not too large values of $S$, ($S \lesssim 10$), the variance for these 
fixed-$S$ distributions rises approximately linearly with $S$ (right panels
of Fig.~\ref{Sdistributions}(b)). This shows that for a given $S$, 
$\Delta X$ is essentially a sum of random contributions, whose number is 
proportional to $S$, the size of the event. This relationship is independent of
$T$ for $\Delta\sigma_s$ (lower right panel), but this holds only for small 
$S$ for $\Delta p$ and $\Delta E$ (upper right panel). It is not clear why this
is. In all cases  the variance tends to saturate as $S$ exceeds 20 (50 for $E$ 
at $T=1.0$).

The above can be made mathematically explicit by writing
$P(\Delta X) = \int_0^\infty P(\Delta X | S)P(S) dS$. At low $T$ if 
$P(S)=(1/\angleb{S})\exp(-S/\angleb{S})$ and the conditional probability 
$ P(\Delta X | S)$ is a Gaussian with variance $\alpha_X S$, then
integration gives  $P(\Delta X) 
= 1/(2\alpha_X\angleb{S})\exp(-2|\Delta X|/\sqrt{2\alpha_X\langle S\rangle})$. 
Linear fits to the $S<10$ data for $T$=0.33 gives for $\alpha_X$ the values
0.32, 1.5$\times10^{-3}$ and 5.8$\times10^{-3}$ for  $E$, $p$ and $\sigma_s$, 
respectively. The decay lengths of the corresponding EDs should be 
$\sqrt{\alpha_X\langle S\rangle/2}$. For $T$=0.33, $\langle S\rangle$=1.83, we 
get values 0.54, 0.012 and 0.023. After normalizing by the r.m.s. fluctuations
as was done in Fig.~\ref{histEventSizes} these become 0.16, 0.15, 0.26, in
reasonable agreement with the decay lengths determined in 
Fig.~\ref{histEventSizes}.

% \footnote{We have also tested the
% reverse hypothesis---that $S$ is 
%controlled by, say, $\Delta^NE$. Taking events with a particular value (or 
%small interval) of $\Delta^NE$ we examine their $S$-distribution, and find case
%not a Gaussian but again an exponential tail, with the same decay length  as
% the full $S$-distribution for that temperature. Thus
%constraining $\Delta^NE$ does not particularly constrain $S$ whereas,
%(as shown already) constraining $S$ constrains $\Delta^NE$.}.

To understand the distributions of $S$ we consider the distributions of 
individual (large) particle displacements $d$ in flow events, shown in
Fig.~\ref{histFlowEventDisps}. As found in
 Refs.~\onlinecite{Schroder/others:2000} and \onlinecite{Vogel/others:2004},
clear exponential tails appear for $d$ larger than about 0.2. The decay lengths
vary surprisingly little over the simulated temperature range, 0.08--0.15.
Most of the variation is in the number of particles in the tail region (the
average number per event). We find it useful to take the length scale 
$\sim 0.1$ suggested by the decay length seriously, and define the 
``active'' particles as those with $d > d_L$=0.1 (independent of $T$ 
for simplicity).

Distributions of $N_A$, the number of active particles,
are shown in the right panel of  Fig.~\ref{histFlowEventDisps}. There is a 
striking shift in both the shape and the mean value from high to low $T$:
A typical event at high $T$ involves most of the system; indeed, larger events 
are more probable than smaller ones. At low $T$ the distribution becomes 
exponential with a mean of order a few tens of particles. In this regime, the
events are relatively localized though not necessarily compact (see the
 example in Fig.~\ref{histFlowEventDisps})--string-like spatial correlation
is present in varying degrees in small and intermediate-sized events, while
the largest events tend to have a more compact structure.
The nature of spatial correlations/structure of flow events has been 
investigated by
different authors. While Schr{\o}der et al. found string-like correlations for
transitions between minima, Appignanesi et al. used the distance matrix
method to identify transitions between so-called 
meta-basins \cite{Appignanesi/others:2006}. Their analysis
highlights events involving a large fraction of the system, which was 
relatively small (150 particles). In fact their data (Fig.~2 of
 Ref.~\onlinecite{Appignanesi/others:2006}) is consistent
with an exponential tail with characteristic size of order 10-20 particles;
the ``democratic events'' would simply be the largest events in the
tail. On the other hand we observe string-like features in smaller events, but
not so much in larger ones. Thus our results encompass both those of 
Schr{\o}der et al. and Appignanesi et al.

\begin{figure}
\epsfig{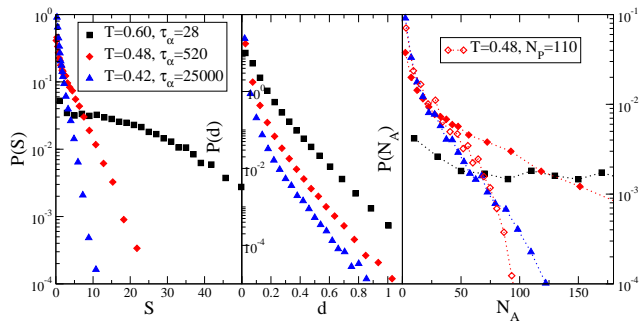}
\caption{\label{distributions3D}  (Color online)Distributions of $S$, $d$ 
(L-particles) 
and $N_A$ for a 3D system at temperatures 
$T$=0.60, 0.48 and 0.42. $P(N_A)$ for a system with $N_P$=110 at $T=0.48$
is also shown.}
\end{figure}

Fig.~\ref{distributions3D} shows data for a 3D system.  
The basic features are similar to the 2D case, and in particular
the $S$ distribution controls the others.
Also shown in the third panel Fig.~\ref{distributions3D} is the distribution
of $N_A$ for a smaller system, $N_P$=110, which is clearly cut off by the 
system size---this could well explain a factor of four difference in relaxation
time between the two sizes that we observe.

The notation $d_L$ is meant to suggest a 
Lindemann-like interpretation. Several authors have 
emphasized the importance of such a length scale in the context of mechanical 
instabilities associated with structural relaxation in 
liquids \cite{Lubchenko/Wolynes:2007,
Chakravarty/Debenedetti/Stillinger:2007, Dudowicz/Freed/Douglas:2005},
also in experiments \cite{Schall/Weitz/Spaepen:2007}. Its
appearance in the present context can be interpreted as
 that events are in a loose sense 
``discretized'' in units of $d_L$: each event involves some number of particles
each of which is displaced some number of $d_L$ units. Exponential distributions
may be interpreted as the statement that there is no preferred number of basic
units (this is analogous to a simple model for polymer size
 distributions, where the  probability to attach a new monomer does not depend
on the current length of the chain \cite{Flory:1953}).

The cross-over to exponential distributions is associated with one
to well-defined localized events, whose mean size decreases with $T$. The 
decrease indicates
increasing material stability. A similar result was found by Fabricius
and Stariolo when perturbing equilibrium configurations and comparing the
corresponding inherent states \cite{Fabricius/Stariolo:2002}. If event sizes 
decrease with
decreasing temperature, this presumably has consequences for relaxation times,
because a given change takes more events, although decreasing size
(in particular $\Delta E$) also suggests that energy barriers for individual 
events also decrease (we will study energy barriers in upcoming work). On the
other hand, when there is a distribution of barriers, there should be 
increased tendency for repeated reverse-transitions. This could be examined
by studying correlations between events, which we have ignored here,
but will address in upcoming work. One way to account for forward-backward
correlations, proposed by Doliwa and Heuer \cite{Doliwa/Heuer:2003b}, involves
grouping minima into larger structures called metabasins and
study transitions between these. Their analysis indicates growing effective
barriers as $T$ decreases, consistent with non-Arrhenius slowing
down of dynamics. Our analysis focuses on the more fundamental units of
dynamics, individual minima-transitions, because these are (in principle) 
unambiguously defined and it is worth understanding their statistics in detail
before attempting to coarse-grain. The analysis 
suggests the somewhat paradoxical result that individual event
barriers decrease as $T$ does. This emphasizes even more the role of 
correlations.

\begin{acknowledgments}
Center for viscous liquid dynamics ``Glass And Time'' is sponsored by The 
Danish National Research Foundation.
\end{acknowledgments}

%\bibliography{Bailey,Roskilde,GlassExperiment,GlassTheorySim,Methods,Miscellaneous,GeneralTheory,LiquidTheory}

\end{document}